\documentclass[superscriptaddress]{revtex4}
\usepackage{amsmath,lscape}
\usepackage{graphicx}
\usepackage{floatflt,epsfig}
\begin{document}

\def\qgp{quark--gluon--plasma}
\def\njl{Nambu--Jona--Lasinio}
\def\cs{chiral symmetry}
\def\ii{\'{\i}}
\def\d{\mbox{d}}

\title{Equation of State for supernova explosion simulations}

\author{D.P.Menezes}
\affiliation{Depto de F\'{\i}sica - CFM - Universidade Federal de Santa
Catarina  Florian\'opolis - SC - CP. 476 - CEP 88.040 - 900 - Brazil}

\author{C. Provid\^encia}
\affiliation{Centro de F\ii sica Te\'orica - Dep. de F\ii sica - 
Universidade de Coimbra - P-3004 - 516 Coimbra - Portugal}

\begin{abstract}
In this work we present a detailed explanation of the construction of an 
appropriate equation of state (EoS) for nuclear astrophysics.
We use a relativistic model in order to obtain an EoS for neutrally charged 
matter that extends from very low to high
densities, from zero temperature to 100 MeV with proton fractions ranging from
0 (no protons) to 0.6 (asymmetric matter with proton excess).
For the achievement of complete convergence, the Sommerfeld approximation is 
used  at low temperatures and the Boltzman distribution for relativistic particles
is used in the calculation of the electron properties at very low densities.
Photons are also incorporated as blackbody radiation. An extension of this EoS
is also presented with the inclusion of strangeness by taking into account 
the $\Sigma^-$ hyperon only. Strangeness fractions range from 0.02 to 0.3.
\end{abstract}

\maketitle

\section{Introduction}

The observed pulsars, also commonly known as neutron stars, are believed to be 
the remnants of type II supernova explosions. These type II supernovae appear at 
the end of the evolution of very massive stars. The core of these stars collapses
to a density around several times nuclear saturation density. A rebound takes 
place and drives a shockwave which expells most of the original mass of the star.
The simulation of supernova explosions and the conditions for them to take place
have been subject of investigation for the last 30 years. Depending on certain
thermodynamical conditions present in the equations of state (EoS), the supernova 
explosion simulation is successful or not \cite{cooperstein}. 
The EoS built for nuclear astrophysics purposes depends on a series of 
thermodynamic properties which are obtained for certain temperatures, densities
and matter composition. Hence, an efficient EoS which is reasonably accurate is
mandatory for a supernova explosion simulation to be successful.

In order to obtain an equation of state (EoS) for low and high
density matter suitable to astropysical applications, the relativistic non 
linear Walecka model (NLWM) 
\cite{walecka,bb} is used. For matter to be neutral, electrons are also included.
For sufficiently high densities the formation of hyperons is energetically 
favored. Normally, the appearance of the strange baryons softens the EoS.
Our formalism is described next with the inclusion of the whole baryonic octet
for the sake of completeness but, in a first step towards a complete desciption
of a supernova explosion, only protons and neutrons are considered. Next 
we incorporate strangeness but restrict ourselves to the inclusion of $\Sigma^-$.
Convergence problems are well known to exist at low temperatures and below certain
densities. Appropriate approximations are then utilized.
Blackbody radiation is considered and, whenever convenient, electrons and positrons
are described separately.
Future prospects for obtaining more sophisticated EoS are discussed.

\section{Hadronic matter equation of state}

A common extension of the NLWM  considers the inclusion of the whole 
baryonic octet ($n$, $p$, $\Lambda$, $\Sigma^+$, $\Sigma^0$, $\Sigma^-$, 
$\Xi^-$,  $\Xi^0$) in the place of the nucleonic sector. In this work we present 
the complete formalism, but numerical calculations were performed with nucleons
only.

The lagrangian density of the NLWM reads:
\begin{equation}
{\cal L}={\cal L}_{B}+{\cal L}_{mesons}+{\cal L}_{leptons},
\label{octetlag}
\end{equation}
where
$$
{\cal L}_B= \sum_B \bar \psi_B \left[\gamma_\mu\left (i\partial^{\mu}
-g_{vB} V^{\mu}- g_{\rho B} {\mathbf t} \cdot {\mathbf b}^\mu \right) 
-(M_B-g_{s B} \phi)\right]\psi_B,
$$
with $\sum_B$ extending over the chosen baryons B,
$$g_{s B}=x_{s B}~ g_s,~~g_{v B}=x_{v B}~ g_v,~~g_{\rho B}=x_{\rho B}~ 
g_{\rho}$$
and $x_{s B}$, $x_{v B}$ and $x_{\rho B}$ are equal to $1$ for the nucleons and
acquire different values in different parametrizations for the other baryons,
$$
{\cal L}_{mesons}=\frac{1}{2}(\partial_{\mu}\phi\partial^{\mu}\phi
-m_s^2 \phi^2) - \frac{1}{3!}\kappa \phi^3 -\frac{1}{4!}\lambda
\phi^4
-\frac{1}{4}\Omega_{\mu\nu}\Omega^{\mu\nu}+\frac{1}{2}
m_v^2 V_{\mu}V^{\mu} 
$$
\begin{equation}
-\frac{1}{4}{\mathbf B}_{\mu\nu}\cdot{\mathbf B}^{\mu\nu}+\frac{1}{2}
m_\rho^2 {\mathbf b}_{\mu}\cdot {\mathbf b}^{\mu},
\end{equation}
where
$\Omega_{\mu\nu}=\partial_{\mu}V_{\nu}-\partial_{\nu}V_{\mu} ,
$
$
{\mathbf B}_{\mu\nu}=\partial_{\mu}{\mathbf b}_{\nu}-\partial_{\nu} 
{\mathbf b}_{\mu}
- g_\rho ({\mathbf b}_\mu \times {\mathbf b}_\nu)
$
and ${\mathbf t}$ is the isospin operator.

In the above lagrangian, neither pions nor kaons
are included because they vanish in the mean field 
approximation which is used in the present work and we do not consider 
the possible contribution of pion and kaon condensates. The leptonic sector
is included as a free fermi gas which does not interact with the hadrons. Its 
lagrangian density reads:

\begin{equation}
{\cal L}_{leptons}=\sum_l \bar \psi_l \left(i \gamma_\mu \partial^{\mu}-
m_l\right)\psi_l.
\end{equation}

In the present work only electrons (and positrons are considered).
In the mean field approximation (MFA) (see \cite{compact,ddpeos}, for 
instance), the meson equations of motion read:
\begin{equation}
\phi_0=- \frac{\kappa}{2 m_s^2} \phi_0^2 
-\frac{\lambda}{6 m_s^2}\phi_0^3 + \sum_B \frac{g_s}{m_s^2} x_{s B}~\rho_{s B},
\label{octphi0}
\end{equation}
\begin{equation}
V_0 = \sum_B \frac{g_v }{m_v^2} x_{v B}~ \rho_B,
\end{equation}
\begin{equation}
b_0 = \sum_B \frac{g_{\rho}}{m_{\rho}^2} x_{\rho B}~  t_{3B}~ \rho_B,
\label{octb0}
\end{equation}
with
\begin{equation}
\rho_B=2 \int\frac{\d^3p}{(2\pi)^3}(f_{B+}-f_{B-}), \quad
\rho=\sum_B \rho_B,
\label{rhob}
\end{equation}
$$
\rho_{s B}= \frac{1}{\pi^2} \int
p^2 \d p \frac{M_B^*}{\sqrt{p^2+{M_B^*}^2}} (f_{B+}+f_{B-}),
$$
with $M_B^*=M_B - g_{s B}~ \phi$, $B\pm$ stands respectively for baryons
and anti-baryons, $t_{3B}$ is the third component of the
baryon isospin,
$E^{\ast}({\mathbf p})=\sqrt{{\mathbf p}^2+{M^*}^2}$ and
\begin{equation}
f_{B\pm}=
{1}/\{1+\exp[(E^{\ast}({\mathbf p}) \mp \nu_B)/T]\}\;,
\end{equation}
where the effective chemical potential is 
\begin{equation}
\nu_B=\mu_B - g_{v B} V_0 - g_{\rho B}~  t_{3 B}~ b_0.
\end{equation}

Within the MFA, the meson fields are taken as classical fields whilst the
baryon fields remain quantum \cite{walecka}. On the other hand, the Dirac
equation, which is the equation of motion for the baryons is not solved
directly but instead used in the calculation of the densities appearing in the
meson equations of motion. The system has then to be solved self-consistently.

At $T=0$, the distribution functions for baryons are replaced by
step functions. In this case equation (\ref{rhob}) becomes simply
$\rho_B={k_{FB}^3}/{3\pi^2}.$
The baryonic energy density  in the mean field approximation reads:
\begin{eqnarray}
{\cal E}_B &=& 2 \sum_B \int \frac{\d^3p}{(2\pi)^3}
\sqrt{{\mathbf p}^2+{M^*}^2} \left(f_{B+}+f_{B-}\right)+
\nonumber \\
&&\frac{m_s^2}{2} \phi_0^2 + \frac{\kappa}{6} \phi_0^3
+\frac{\lambda}{24}\phi_0^4
+\frac{m_v^2}{2} V_0^2 + \frac{\xi g_v^4}{8} V_0^4 
+\frac{m_{\rho}^2}{2} b_0^2
\label{ener}
\end{eqnarray}
and the related pressure becomes
\begin{eqnarray}
P_B&=&\frac{1}{3 \pi^2} \sum_{B}
\int \d p \frac{{\mathbf p}^4}{\sqrt{{\mathbf p}^2+{M^*}^2}} 
\left( f_{B+} + f_{B-}\right)
\nonumber \\
&&-\frac{m_s^2}{2} \phi_0^2 -\frac{\kappa\phi_0^3}{6} -
\frac{\lambda\phi_0^4}{24}
+\frac{m_v^2}{2} V_0^2 +\frac{\xi g_v^4 V_0^4}{24}
+\frac{m_{\rho}^2}{2} b_0^2.
\label{press}
\end{eqnarray}
The entropy of the baryons are taken as
\begin{eqnarray}
{\cal S}_B &=& -2 \sum_B \int \frac{\d^3p}{(2\pi)^3}
\left[f_{B+} log (f_{B+}) + (1-f_{B+}) log (1-f_{B+}) \right.
\nonumber \\
&& + \left. f_{B-} log (f_{B-}) + (1-f_{B-}) log (1-f_{B-}) \right]
\label{entropy}
\end{eqnarray}
and hence the free energy reads
\begin{equation}
{\cal F}_B={\cal E}_B - T {\cal S}_B .
\end{equation}

Notice again that the above expressions were obtained for finite temperature, 
but they can be easily modified for $T=0$. Whenever T=0, no anti-particles are 
present.

At very low T ($-0.4 \leq log(T) < -0.1$ MeV) there are well known convergence 
problems due to the
distribution functions and in this case we use the Sommerfeld
approximation for the baryons \cite{sommer}. The effective chemical 
potentials, in particular, read

\begin{equation}
\nu_i=\epsilon_{Fi}-\frac{\pi^2}{6} \, T^2\, 
\frac{\left(k_{Fi}^2+\epsilon_{Fi}^2\right)}{k_{Fi}\epsilon_{Fi}},
\quad i=p,n.
\end{equation}


For the net electron density we have
\begin{equation}
\rho_e=2 \int\frac{\d^3p}{(2\pi)^3}(f_{e^-}-f_{e^+}), \qquad  
\label{rhol}
\end{equation}
where  the distribution functions for the particles ($e^-$) and
antiparticles ($e^+$) are given by
\begin{equation}
f_{e^{\mp}}=1/({1+\exp[(\epsilon\mp\mu_e)/T]})\;,
\label{distf}
\end{equation}
with $\mu_e$  as the chemical potential. In order to ensure charge neutrality,
electron and proton densities have to be equal, i.e.,
\begin{equation}
\rho_e=\rho_p.
\label{rhol2}
\end{equation}
Next we always distinguish between
electrons ($e^-$) and positrons ($e^+$) and when both particles and antiparticles 
are considered we refer to the related quantity with the index $e$.
At $T=0$, the distribution functions for the leptons are also replaced by
step functions and no positrons are left.
In this case equation (\ref{rhol}) becomes simply
$\rho_e={k_{Fe}^3}/{3\pi^2}.$
The thermodynamic quantities read

\begin{equation}
{\cal E}_e = 
2 \int \frac{d^3p}{{(2 \pi)}^3} \sqrt{{\mathbf p}^2+m_e^2}
(f_{e^-} + f_{e^+}),
\end{equation}
\begin{equation}
{\cal E}_{e^-} = 
2 \int \frac{d^3p}{{(2 \pi)}^3} \sqrt{{\mathbf p}^2+m_e^2} ~
f_{e^-}, \quad
{\cal E}_{e^+} = 
2 \int \frac{d^3p}{{(2 \pi)}^3} \sqrt{{\mathbf p}^2+m_e^2} ~
f_{e^+},
\end{equation}

\begin{equation}
P_e=
\frac{1}{3 \pi^2} \int \frac{{\mathbf p^4} dp}
{\sqrt{{\mathbf p}^2+m_e^2}} (f_{e^-} + f_{e^+}),
\end{equation}
\begin{equation}
P_{e^-}=
\frac{1}{3 \pi^2} \int \frac{{\mathbf p^4} dp}
{\sqrt{{\mathbf p}^2+m_e^2}} ~f_{e^-}, \quad
P_{e^+} = P_e - P_{e^-}
\end{equation}

\begin{equation}
{\cal S}_e=\frac{{\cal E}_e + P_e - \mu_e \rho_e}{T},
\quad
{\cal S}_{e^-}=\frac{{\cal E}_{e^-} + P_{e^-} - \mu_e \rho_{e^-}}{T},
\quad
{\cal S}_{e^+}=\frac{{\cal E}_{e^+} + P_{e^+} + \mu_e \rho_{e^+}}{T},
\end{equation}

\begin{equation}
{\cal F}_e={\cal E}_e - T {\cal S}_e , \quad
{\cal F}_{e^-}={\cal E}_{e^-} - T {\cal S}_{e^-} , \quad
{\cal F}_{e^+}={\cal E}_{e^+} - T {\cal S}_{e^+} ,
\end{equation}

The particle fraction is defined as
$y_i=\rho_i/\rho$, where $i=p,n,e^-,e^+$, and $\rho$ is the total baryonic density.

At very low densities a Boltzman distribution for relativistic electrons and
positrons is necessary \cite{boltz}. The low density limit depends on the
temperature and is numerically chosen such that eqs. (\ref{rhol}) and
(\ref{rhol2}) are equal within a $10^{-6}$ precision. If the difference is
larger than this limit, eq. (\ref{rhol2}) is chosen and the corresponding 
chemical potentials are
\begin{equation}
\nu_e=m_e+log\left[ \frac{\rho_e}{g}\left(\frac{2\pi}{T m_e}
\right)^{3/2}\right], 
\end{equation}
with $g=2$ defined as the spin multiplicity, Moreover,
\begin{equation}
\rho_e=(e^{\mu_e/T}-e^{-\mu_e/T})\frac{I_1}{\pi^2},
\end{equation}
or analogously,
\begin{equation}
\mu_e=T \log[\frac{z}{2}+\sqrt{\frac{z^2}{4}+1}], \qquad z=\pi^2\rho_e/I_1.
\end{equation}
The energy density and pressure become
\begin{equation}
{\cal E}_e=(e^{\mu_e/T}-e^{-\mu_e/T})\frac{I_2}{\pi^2},
\end{equation}
\begin{equation}
P_e=(e^{\mu_e/T}-e^{-\mu_e/T})\frac{(I_2- m_e^2 I_0)}{3 \pi^2},
\end{equation}
where
$\beta=1/T$, $I_0=\frac{m_e}{\beta}K_1(m_e\beta)$,
$I_1=-\frac{dI_0}{d\beta}=\frac{m_e}{\beta^2}K_1(y) 
-\frac{m_e^2}{\beta}\frac{dK_1}{d y}$,
$I_2=-\frac{dI_1}{d\beta}=\frac{2m_e}{\beta^3}K_1(y) -2\frac{m_e^2}{\beta^2}
\frac{dK_1}{dy}+\frac{m_e^3}{\beta}\frac{d^2K_1}{d y^2}$,
with
$y=m_e\beta$, $K_i$ are modified Bessel functions and
$K'_\nu(x)=-\frac{1}{2}\left(K_{\nu-1}(x)+K_{\nu+1}(x)\right)$.


The photons are taken into account via blackbody radiation and the main 
expressions are

\begin{equation}
P_\gamma=\frac{\pi^2 T^4}{45}, \quad {\cal S}_\gamma=\frac{4 P_\gamma}{T},
\quad  {\cal E}_\gamma=3 P_\gamma, \quad 
{\cal F}_\gamma= {\cal E}_\gamma-T {\cal S}_\gamma . 
\end{equation}


For the hadron phase we have used the GM3 parametrization proposed by
Glendenning and Moszkowski \cite{gm91}, corresponding to an effective mass 
$M^*=0.78\, M$ and incompressibility $K=240$ MeV at the saturation density 
$\rho_0=0.153$ fm$^{-3}$. The coupling constants are
$\left(\frac{g_s}{m_s}\right)^2=9.927$,
$\left(\frac{g_v}{m_v}\right)^2=4.82$,
$\left(\frac{g_\rho}{m_\rho}\right)^2=4.79$,
$\kappa=0.017318 gs^3$, $\lambda=-0.014526 gs^4$.

In our codes the inputs are the temperature, proton fraction and baryonic 
density. The grids for these quantities are
$-0.4 \leq log (T) \leq 2$ (MeV) with mesh intervals of 0.1,\\ 
$0 \leq y_p \leq 0.6$ with mesh intervals of 0.02, \\
$3 \leq log (\rho) \leq 15.7$ (g/cm$^3$). 

In the output we have
$y_p$, $y_n$, $y_{e^-}$, $y_{e^+}$, $y_e$, \\
$\mu_p-M$ (MeV), $\mu_n-M$ (MeV), $\mu_e-m_e$ (MeV), $-\mu_e+m_e$ (MeV),\\
${\cal E}_B$ (erg/g), ${\cal E}_{e^-}$ (erg/g), ${\cal E}_{e^+}$ (erg/g), 
${\cal E}_\gamma$ (erg/g), ${\cal E}_B + {\cal E}_e + {\cal E}_\gamma$ (erg/g),\\
${\cal S}_B$ ($k_B$/baryon), ${\cal S}_{e^-}$ ($k_B$/baryon), 
${\cal S}_{e^+}$ ($k_B$/baryon), ${\cal S}_\gamma$ ($k_B$/baryon), 
${\cal S}_B + {\cal S}_e + {\cal S}_\gamma$ ($k_B$/baryon),\\
$P_B$ (dyne/cm$^2$), $P_{e^-}$ (dyne/cm$^2$), $P_{e^+}$ (dyne/cm$^2$),
$P_\gamma$ (dyne/cm$^2$), $P_B + P_e + P_\gamma$ (dyne/cm$^2$),\\
${\cal F}_B+{\cal F}_e+{\cal F}_\gamma$ (MeV/fm$^3$).

\section{Threshold density for matter with strangeness}

To include strangeness in the EoS, 
$\Sigma^-$ was first chosen because in $\beta$-
equilibrium matter at zero temperature and with the GM3 parametrization,
its onset appears at lower densities 
than the onset of the least massive hyperon, the $\Lambda$. Depending on the 
parametrization chosen for the NLWM 
and on the hyperon-meson coupling constants, this trend may change at higher
temperatures and hence, in a future work $\Sigma^-$ and
$\Lambda$ should be included simultaneously.

In compact stars, stellar matter is in chemical equilibrium, which means that
$$\mu_{\Sigma^0}=\mu_{\Xi^0}=\mu_{\Lambda}=\mu_n, \quad
\mu_{\Sigma^-}=\mu_{\Xi^-}=\mu_n+\mu_e, \quad
\mu_{\Sigma^+}=\mu_p=\mu_n-\mu_e.$$

In an explosive enviroment like the one existing in a supernova, chemical 
equilibrium is not supposed to be enforced. However, we  
consider that the time during which the supernova explosion occurs is
much longer than the characteristic time of the weak interaction in such a way 
that the strangeness fraction is expected to be finite.

In order to build an EoS containing strangeness and appropriate for a 
supernova simulation we define for each energy density, temperature and proton 
fraction a  threshold density above which a given fraction of strangeness, $y_s$,
is allowed to exist.
We  determine the threshold density from the condition of 
$\beta$-equilibrium for the $\Sigma^-$, which is imposed through the two 
independent chemical potentials ($\mu_n$ and $\mu_e$).
In this case, at $T=0$, the corresponding effective chemical potential
and density are
$$\nu_{\Sigma^-}=\sqrt{k_{F{\Sigma}}^2+{M^*}^2_{\Sigma}}=
\mu_{\Sigma^-}-g_{v\Sigma} V_0+g_{\rho\Sigma} b_0,$$
and
$$\rho_{\Sigma}=\frac{k_{F{\Sigma}}^2}{3\pi^2}.$$
If the condition
$$\frac{\rho_{\Sigma}}{\rho}\ge y_s,$$
is satisfied, the appearence of the strangeness fraction
$y_s$ in the EoS is allowed. For $y_s>0$, we define 
$$\rho=\rho_n+\rho_p+\rho_{\Sigma}$$
with
$$\rho_p=y_p(1-y_s)\rho, \qquad \rho_n=(1-y_p)(1-y_s)\rho, \qquad 
\rho_\Sigma=y_s\rho.$$
For charge neutrality,
\begin{equation}
\rho_e =\rho_p-\rho_{\Sigma^-}
\end{equation}
is required.

In order to fix the meson-hyperon coupling constants we have used the prescription
 given in \cite{gm91,Glen00}, where the hyperon coupling constants are 
constrained by the binding of the $\Lambda$ hyperon in nuclear matter, 
hypernuclear levels and neutron star masses
($x_{s \Sigma}=0.7$ and $x_{v \Sigma}=x_{\rho \Sigma}=0.783$) and assumed that 
the couplings to the $\Sigma$ are equal to those of the $\Lambda$ hyperon.

\section{Results and Future prospects}

A comprehensive test of the thermodynamic accuracy and consistency of our EoS,
as described in \cite{timmes}, mainly when strangeness is introduced,
should be performed.

A non-homogeneous phase known as pasta phase should be considered at low 
densities. This non-homogeneous configuration  made out of spheres, rods, 
bubbles or other more exotic structures, have been extensively used recently 
\cite{pasta,pasta1,maru}. These strucutres may change the neutrino opacity in 
supernova matter and influence neutron star quakes and pulsar glitches.
We can obtain the pasta phase by building the binodal section, and therefore 
obtaining the chemical potentials and densities of the gas and liquid phase in 
equilibrium. A very crude approximation would be to forget Coulomb interaction 
and take zero thickness nuclei. We can consider the matter made of liquid 
dropplets in a gas introducing two parameters: the radius of the Wigner-Seitz 
cell and the radius of the nucleus (equal for protons and neutrons). One of 
the parameters would be fixed imposing a given particle density and the other 
by minimizing the free energy. These results can be improved by including the 
Coulomb contribution and the surface energy by hand. A Thomas-Fermi
calculation can then
be used to obtain the pasta phase with all fields introduced in a consistent
way and the surface energy calculated from the derivatives of the fields.

$\alpha$-particles can be easily incorporated in the EoS as proposed in 
\cite{lattimer}. Once the pasta phase 
and the $\alpha$-particles are included, the EoS should then be compared with
reference \cite{shen}. We are already aware of some important differences.
The parametrization used in \cite{shen}, known as TM1 \cite{tm1} reproduces 
ground state properties of stable and unstable nuclei. 
Nevertheless, this parametrization has proven not to be adequate in the
description of neutron star matter because it breaks down, giving  rise to negative baryon effective
masses at  densities exisiting inside a neutron star (approximately 6 times the nuclear saturation 
density) when hyperons are incorporated into the EoS 
\cite{nossos}. For this reason, we usually choose one of the parametrizations
introduced by Glendenning and collaborators \cite{Glen00}, which give a 
higher nucleon effective mass at the nuclear matter saturation density and, 
for this reason, avoids the problem of the baryon negative masses.
Moreover, according to \cite{shen}, the EoS with inhomogeneties has a critical 
temperature $T \simeq 15$ MeV above which matter is uniform. This number 
certainly depends on the choice of the parameters.  Based on our recent works, 
we would expect a smaller value for the critical temperature since for nuclear 
matter with no electrons (no Coulomb interaction and surface tension) the 
critical temperature occurs for symmetric matter just above 15 MeV. The high 
value obtained in \cite{shen}  may be due to the way the density distributions 
are parametrized which give rise to very stiff surfaces for the
  droplets. In \cite{pasta1} a critical temperature of $\sim 5$ MeV was 
obtained for $y_p=0.3$ matter and  $\sim 6$ MeV was obtained for $y_p=0.5$. 
One of our recent studies on the dynamical instabilities of npe matter also 
predicts 
lower critical temperatures, more according to the results of \cite{pasta1}. 
We do not know if the differences on the EoS due to the use of different 
parametrizations is more important or of the order of the magnitude of the 
changes included due to the explicit inclusion of a non-homogeneous phase. 
This should be studied. 

\begin{figure}[b]
\begin{center}
\begin{tabular}{cc}
$y_p$=0.5 & \epsfig{file=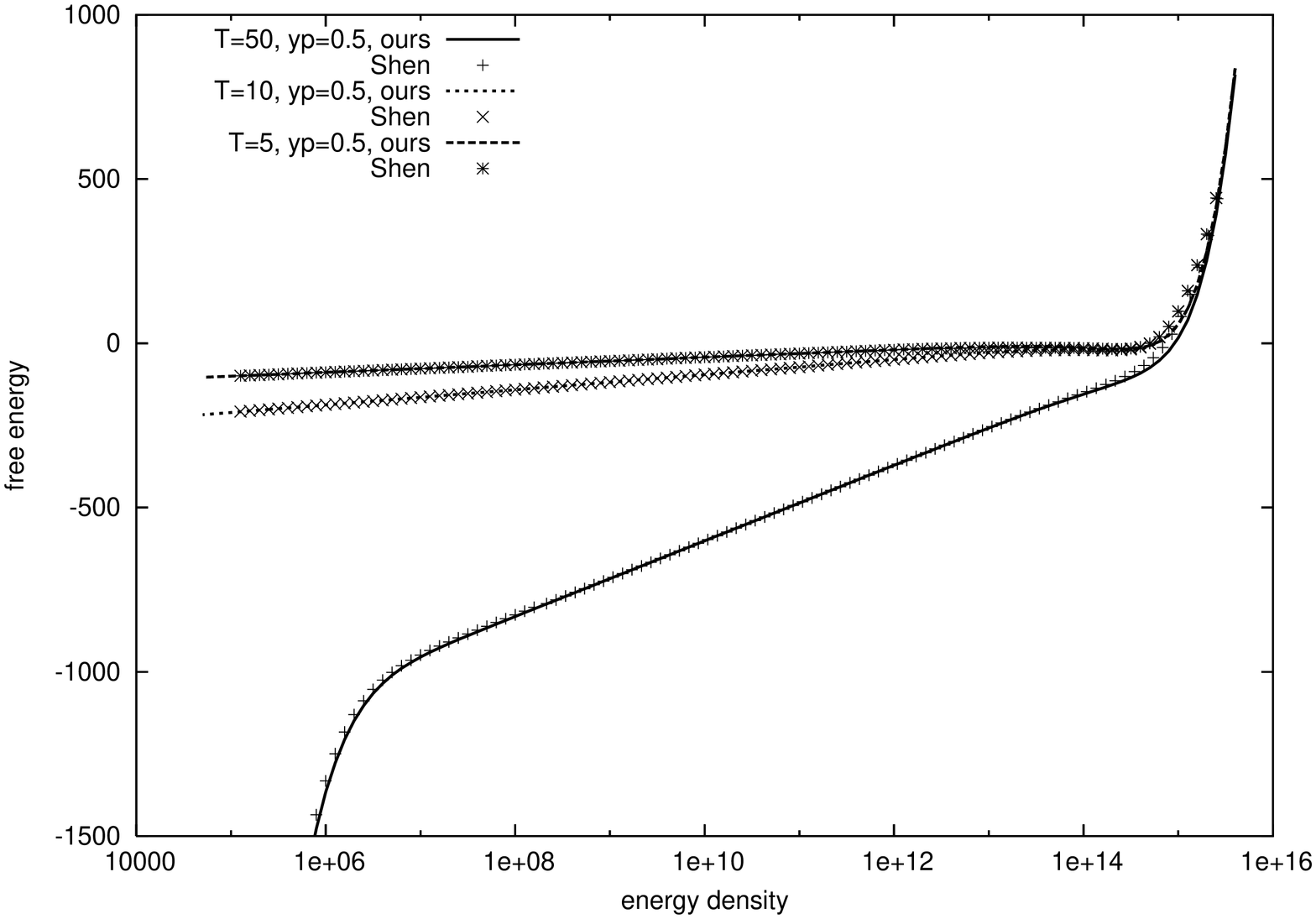,width=7.cm,angle=-90}\\
$y_p$=0.3 & \epsfig{file=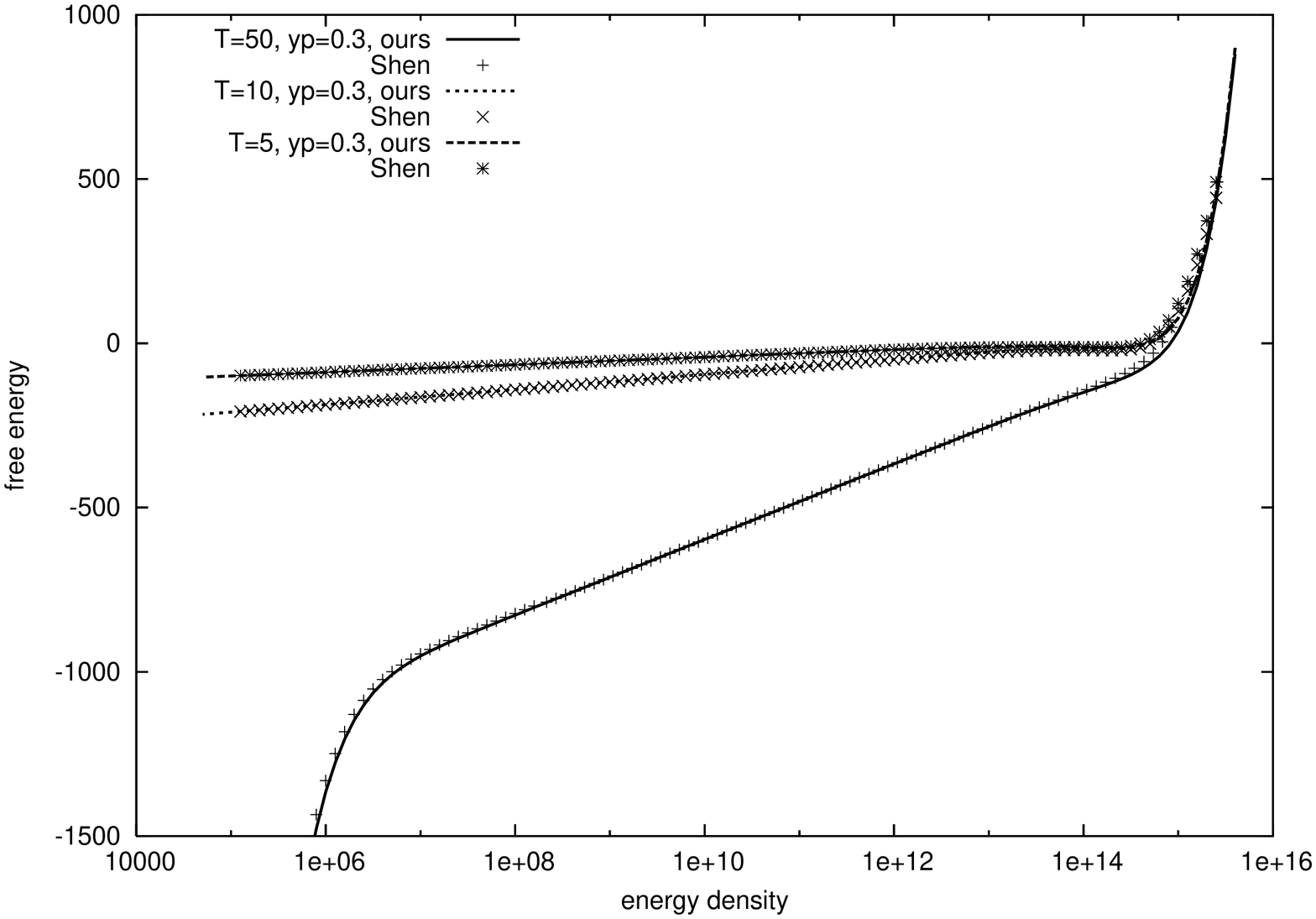,width=7.cm,angle=-90}\\
$y_p$=0.01 & \epsfig{file=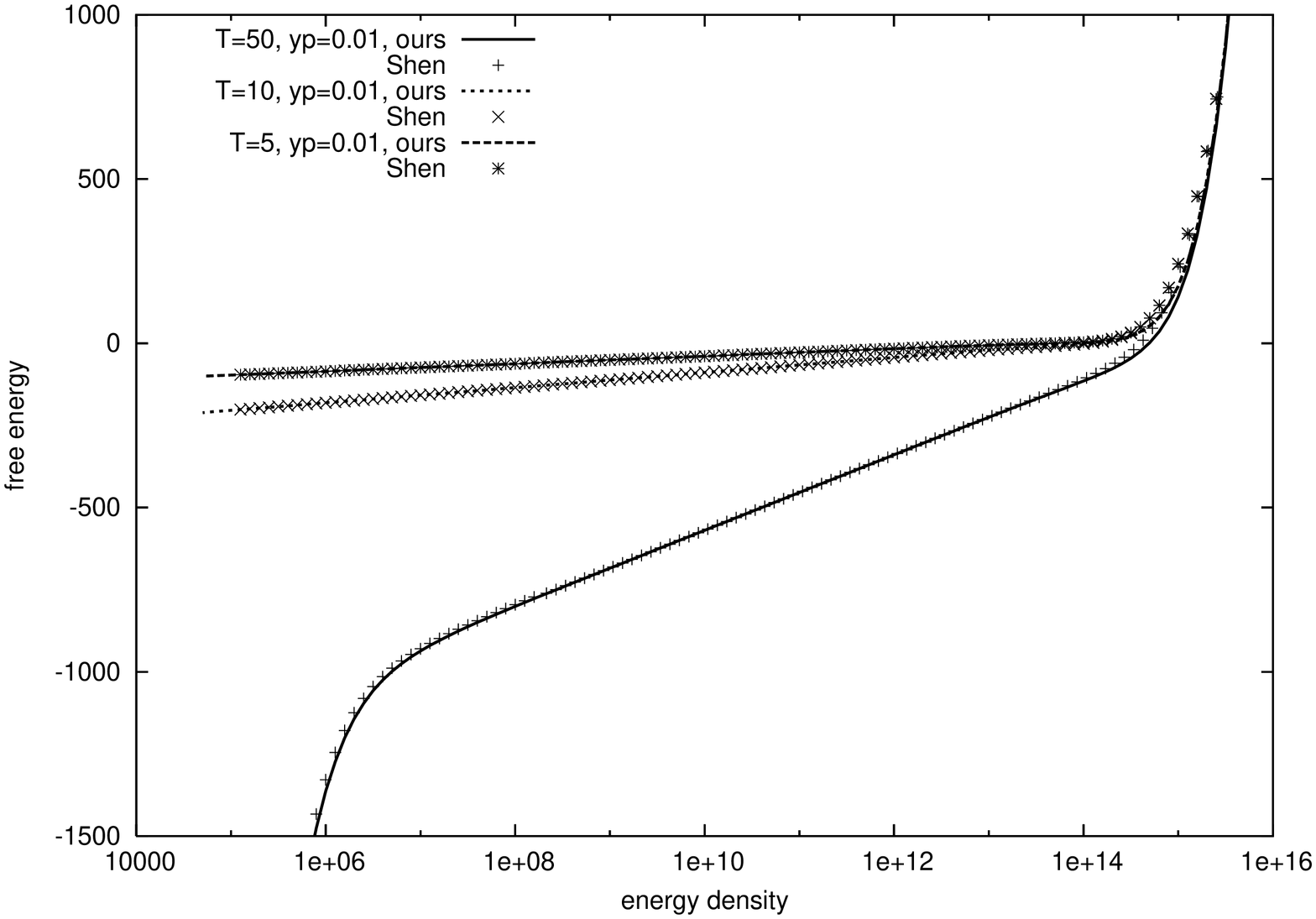,width=7.cm,angle=-90}\\
\end{tabular}
\end{center}
\caption{Free energy (MeV/fm$^3$) as function of the energy density
  (g/cm$^3$) for different temperatures (MeV).}
\label{freeener}
\end{figure}

\begin{figure}[b]
\begin{center}
\begin{tabular}{cc}
$y_p$=0.5 & \epsfig{file=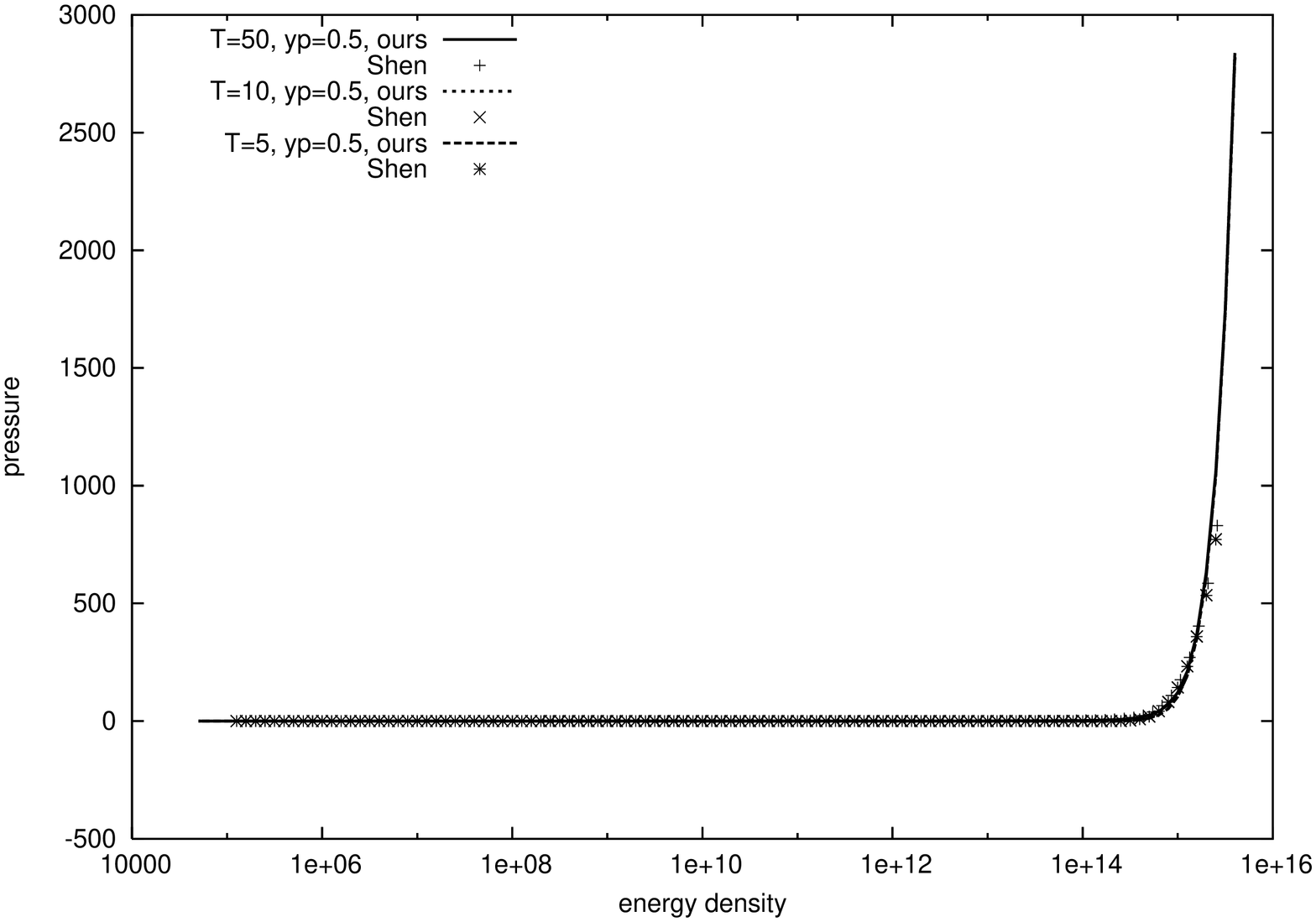,width=7cm,angle=-90}\\
$y_p$=0.3 & \epsfig{file=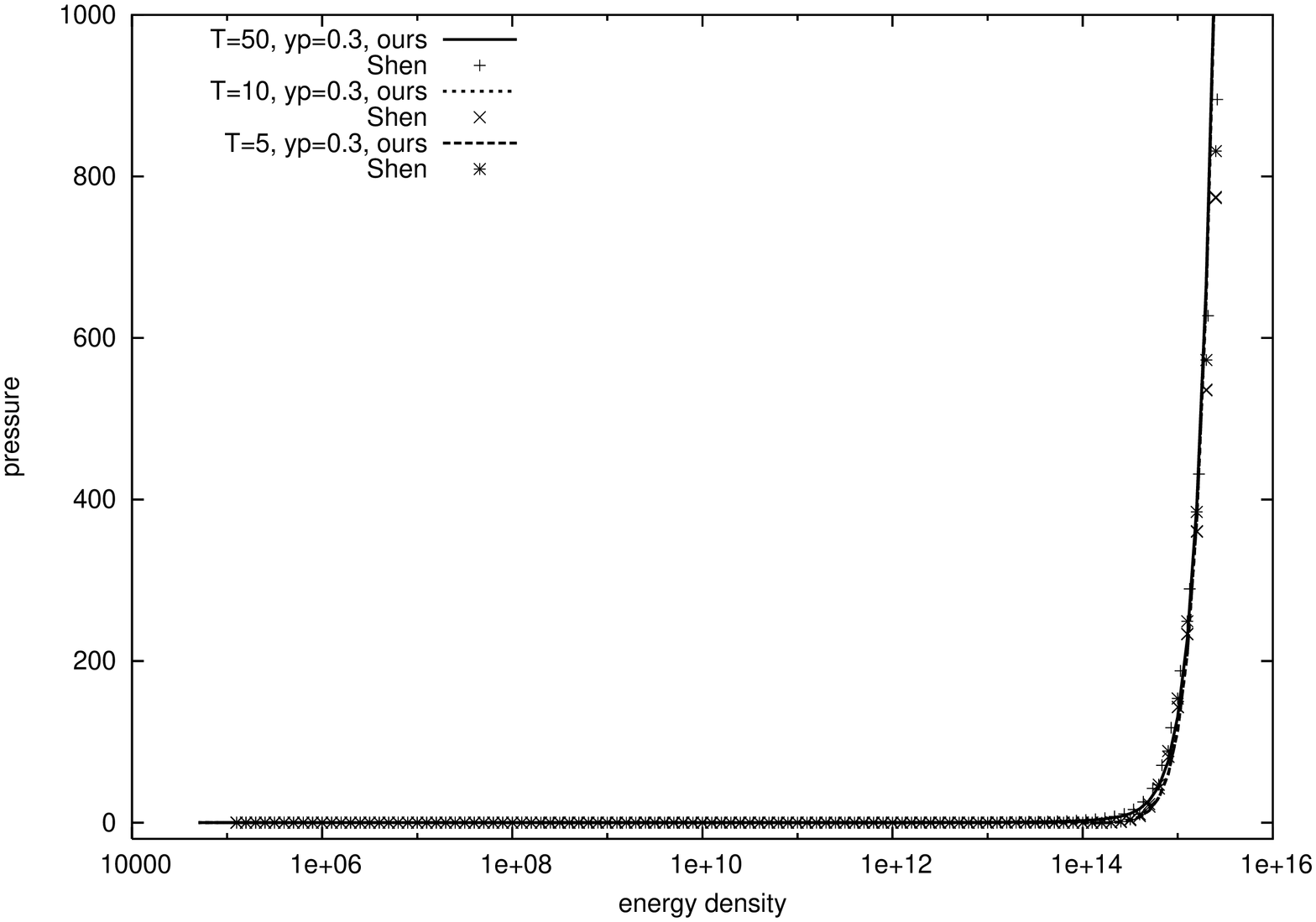,width=7cm,angle=-90}\\
$y_p$=0.01 & \epsfig{file=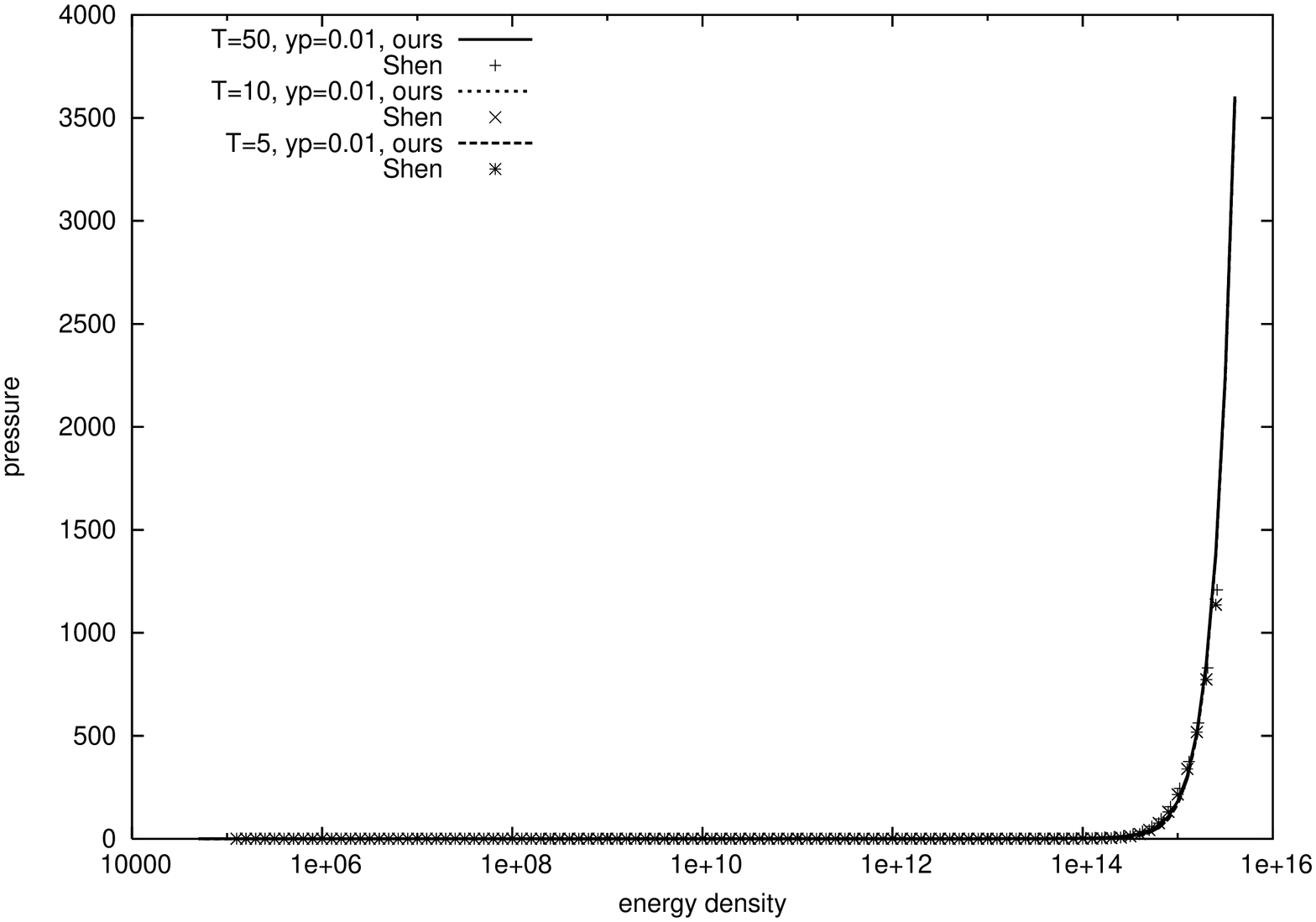,width=7cm,angle=-90}\\
\end{tabular}
\end{center}
\caption{Pressure (MeV/fm$^3$) as function of the energy density
  (g/cm$^3$) for different temperatures (MeV).}
\label{figpress}
\end{figure}

\begin{figure}[b]
\begin{center}
\begin{tabular}{cc}
$y_p$=0.5 & \epsfig{file=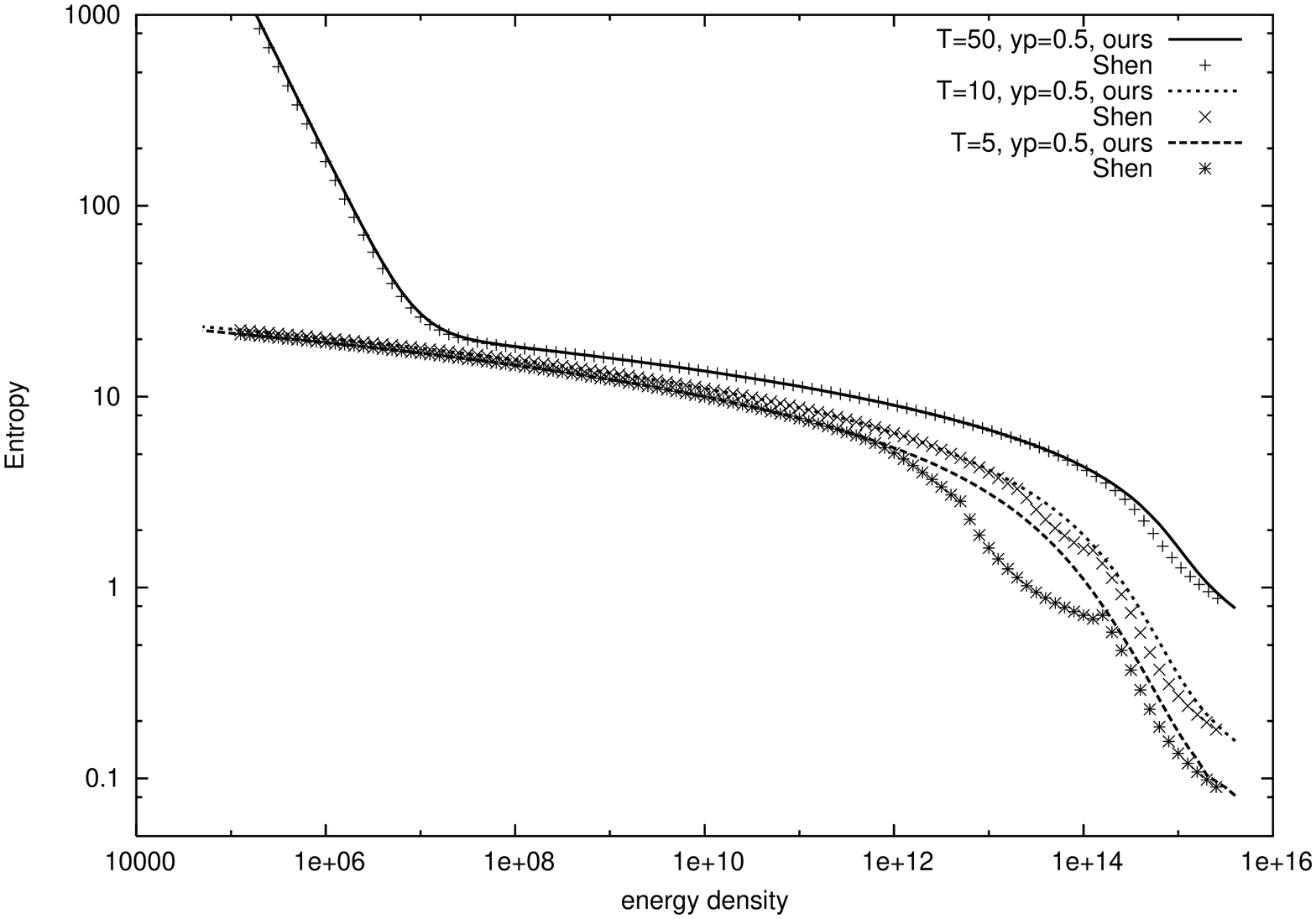,width=7cm,angle=-90}\\
$y_p$=0.3 & \epsfig{file=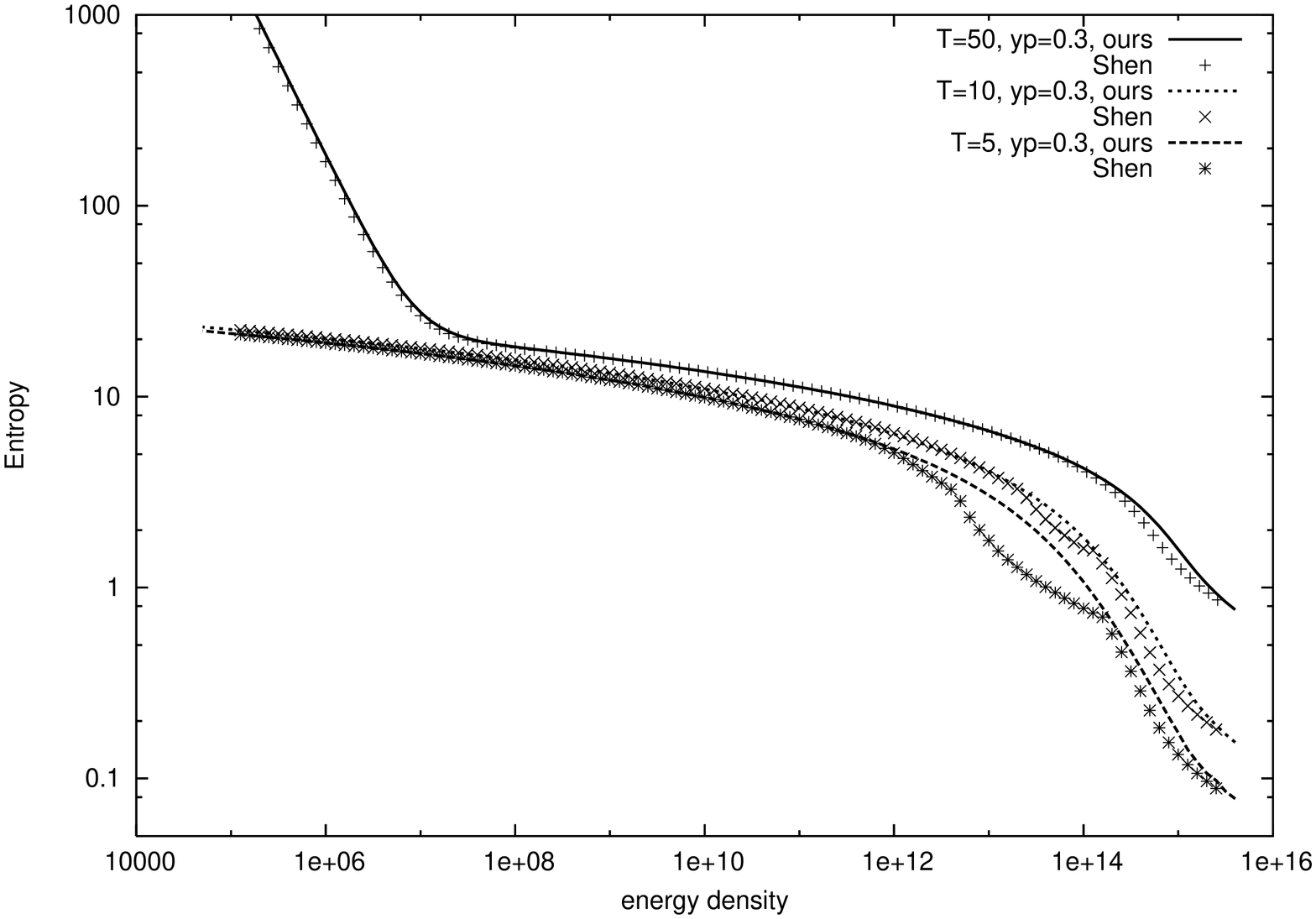,width=7cm,angle=-90}\\
$y_p$=0.01 & \epsfig{file=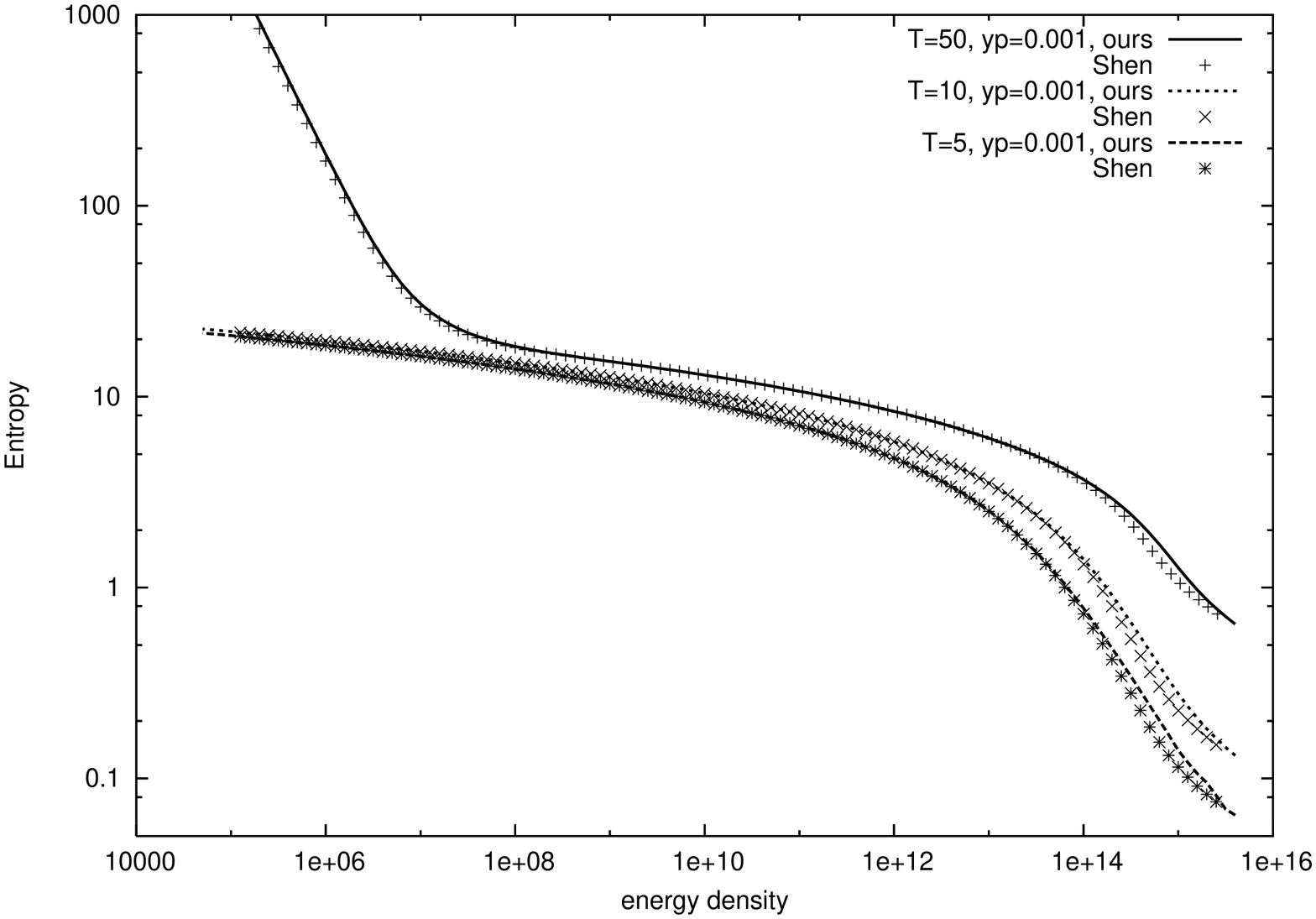,width=7cm,angle=-90}\\
\end{tabular}
\end{center}
\caption{Entropy per baryon as function of the energy density
  (g/cm$^3$) for different temperatures (MeV).}
\label{entr}
\end{figure}

In Fig. \ref{freeener} we compare our results for the 
free energy obtained at three different temperatures and three different proton
fractions with the results of \cite{shen}. One can see that results deviate
sligtly at higher densities. In Fig. \ref{figpress} we plot the pressure for
the same temperatures and proton fractions as in Fig \ref{freeener}. Again the
results are very similar. In Fig. \ref{entr} we plot, once more, for the same
temperatures and proton fractions, the entropy.  The differences are more
pronounced. While at higher temperatures (50 MeV) the curves are very similar
for all proton fraction, for lower temperatures the curves are identical only
for neutron matter (very low proton fraction). One should notice, however,
that the trends of the curves are the same. 

\begin{figure}[b]
\begin{center}
 \epsfig{file=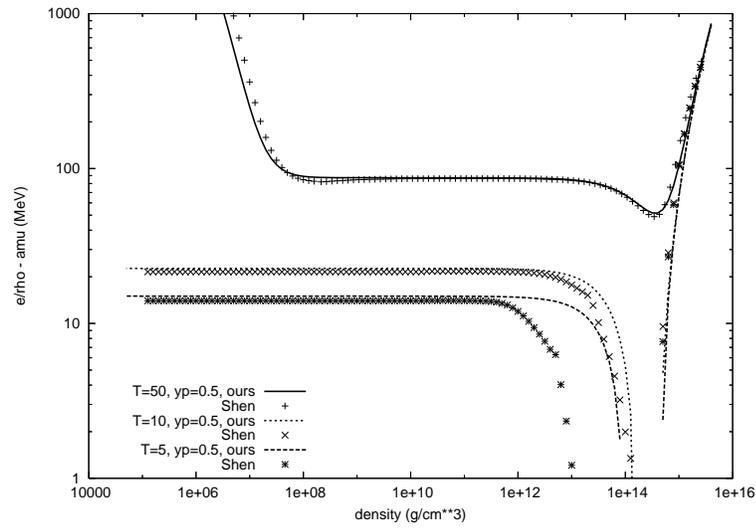,width=7cm,angle=-90}\\
\end{center}
\caption{Internal energy of symmetric matter for different temperatures.}
\label{internal}
\end{figure}

\begin{figure}[b]
\begin{center}
 \epsfig{file=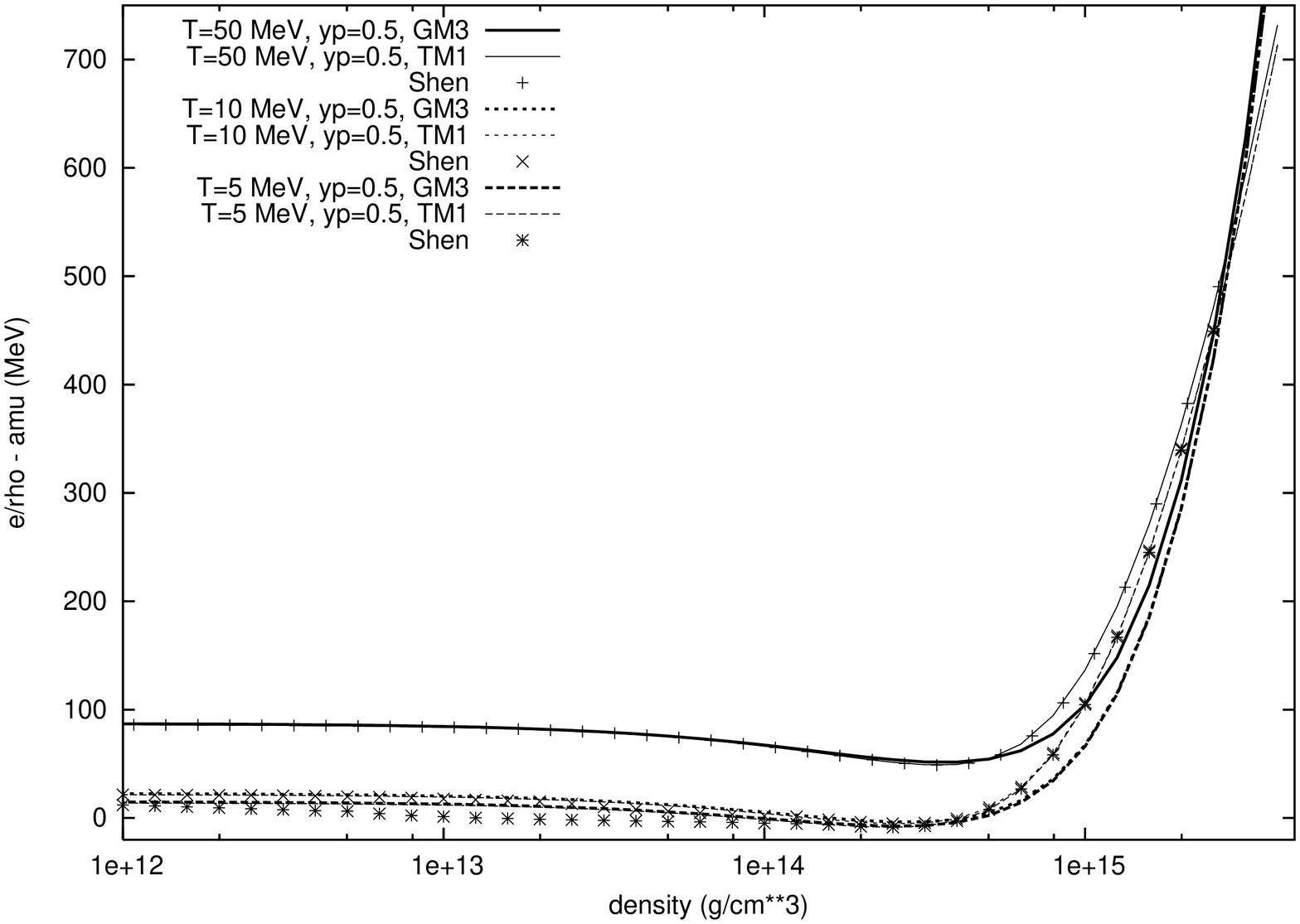,width=7cm,angle=-90}\\
\end{center}
\caption{Internal energy of symmetric matter for different temperatures for homogeneous matter
  within GM3 and TM1 parametrizations, and the EoS of \cite{shen}.}
\label{internal1}
\end{figure}

Finally, we comment on the definition of the internal energy: it is equal to 
the nucleon mass for zero density at T=0 MeV. For finite temperature this is 
no longer true because of the presence of nucleons and antinucleons. 
We have defined the internal energy as the energy density per number density 
in erg/g and in \cite{shen} the internal energy is given by the 
energy density per number density minus the atomic mass unit. As one can see
in Fig \ref{internal}, both results are in accordance once the same definition
is used.
A more clear comparison is done in Fig.  \ref{internal1}  where the internal 
energy for homogeneous matter within TM1 is also shown, and compared with the  
internal energy obtained with the GM3 parametrization and the EoS of 
\cite{shen}, also with TM1 but with the non-homogeneous phase included.

Finally, as a second step in a more refined EoS with strangeness the 
$\Lambda$ hyperons
and later the whole octect and muons should be included.

\section*{ACKNOWLEDGMENTS}

This work was partially supported by CNPq(Brazil),
CAPES(Brazil)/GRICES (Portugal) under project 100/03 and FEDER/FCT (Portugal)
under the projects POCTI/FP/63419/2005 and POCTI/FP/63918/2005.

\newpage


\begin{thebibliography}{99}    

\bibitem{cooperstein} J. Cooperstein, H.A. Bethe and G.E. Brown, Nucl. Phys.
{\bf A 429}, 527 (1984).

\bibitem{walecka} B.D. Serot and J.D. Walecka, Adv. Nucl. Phys. {\bf 16}, 1 
(1986).

\bibitem{bb} J. Boguta and A. R. Bodmer, Nucl. Phys. {\bf A292}, 413 (1977).

\bibitem{compact} D.P. Menezes and C. Provid\^encia, Phys. Rev. {\bf C 68},
035804 (2003).

\bibitem{ddpeos} S.S. Avancini and D.P. Menezes, Phys. Rev. C 74, 015201 (2006).

\bibitem{sommer} N. Ashcroft, N.D. Mermim, Solid State Physics, Saunders
  College Publishing, Orlando, 1976 

\bibitem{boltz} L. D. Landau and E. M. Lifshitz, Statistical Physics, Pergamon Press, 1959.

\bibitem{gm91}  N. K. Glendenning and S. Moszkowski, Phys. Rev. Lett. 
{\bf 67}, 2414 (1991).

\bibitem{Glen00} N. K. Glendenning, Compact Stars, Springer-Verlag, New-York, 
2000.

\bibitem{timmes} F.X. Timmes and D. Arnett, Astrophys. J. Suppl. {\bf 125}, 277
(1999).

\bibitem{lattimer} J.M. Lattimer and F.D. Swesty, Nucl. Phys. {\bf 535}, 331
(1991).

\bibitem{pasta}  D. G. Ravenhall, C. J. Pethick, and J. R. Wilson,
  Phys. Rev. Lett. {\bf 50}, 2066 (1983); M. Hashimoto, H. Seki, and
  M. Yamada, Prog. Theor. Phys.{\bf 71}, 320 (1984).

\bibitem{pasta1}  G. Watanabe, K. Sato, K. Yasuoka, and T. Ebisuzaki,
  Phys. Rev. C 69, 055805 (2004); G. Watanabe, T. Maruyama, K. Sato, K. Yasuoka, and T. Ebisuzaki, Phys.
Rev. Lett. 94, 031101 (2005).

\bibitem{maru} T. Maruyama, T. Tatsumi, D. N. Voskresensky, T. Tanigawa, and S. 
Chiba, Phys. Rev. C 72, 015802 (2005).

\bibitem{shen} H. Shen, H. Toki, K. Oyamatsu, K. Sumiyoshi, Nucl. Phys. {\bf A
  637}, 435 (1998); H. Shen, H. Toki, K. Oyamatsu, K. Sumiyoshi, {\it User Notes
for Relativistic EoS Tables}.

\bibitem{tm1} K. Sumiyoshi, H. Kuwabara, H. Toki, Nucl. Phys. {\bf A
581}, 725 (1995).

\bibitem{nossos} D.P. Menezes and C. Provid\^encia, Phys. Rev. {\bf C 68}, 
035804 (2003); A.M.S. Santos and D.P. Menezes, Phys. Rev. {\bf C 69}, 045803 
(2004).

\end{thebibliography}
\end{document}